\documentclass[a4paper]{jpconf}

\usepackage{amssymb}
\usepackage{amsmath}
\usepackage{amscd}
\usepackage{latexsym}

\usepackage{float}

\usepackage{cite}

\usepackage{graphicx}
\usepackage{subfigure}

\newcommand{\be}{\begin{equation}}
\newcommand{\ee}{\end{equation}}

\newcommand{\bt}{\beta}
\newcommand{\vp}{\varphi}
\newcommand{\ep}{\varepsilon}
\newcommand{\al}{\alpha}
\newcommand{\ra}{\rightarrow}

\newcommand{\cD}{{\cal D}}
\newcommand{\cP}{{\cal P}}
\newcommand{\cH}{{\cal H}}

\newcommand{\cB}{{\cal B}}
\newcommand{\rgl}{\rangle}

\begin{document}

\title{Entanglement production by evolution operator}

\author{V.I. Yukalov$^1$ and E.P. Yukalova$^2$}

\address{
$^1$Bogolubov Laboratory of Theoretical Physics,
Joint Institute for Nuclear Research, \\ Dubna 141980, Russia}

\address{
$^2$Laboratory of Information Technologies,
Joint Institute for Nuclear Research, \\ Dubna 141980, Russia}

\ead{yukalov@theor.jinr.ru}

\begin{abstract}
Entanglement production, generated by an evolution operator, is 
considered. The measure of entanglement production, introduced earlier by one of 
the authors, is employed. As an illustration, a two-qubit register is studied and 
the corresponding measure of evolutional entanglement production is calculated. 
Such two-qubit registers can be realized by atomic systems in a double well or
by trapped atoms with two coherent modes. 
\end{abstract}

\section{Entangled and disentangled states}

Entanglement is a pivotal quantum property that can give a principal advantage
for the use of quantum processes, as compared to classical ones, in such fields 
as quantum information processing, quantum computing, and the functioning of 
artificial quantum intelligence 
\cite{Williams_1,Nielsen_2,Vedral_3,Keyl_4,YS_5,Wilde_6,YS_7}. 

The states, represented by wave functions or state vectors, are distinguished
onto non-entangled and entangled, as explained below. 

One considers a composite system characterized by a Hilbert space
\be
\label{1}
\cH = \bigotimes_{i=1}^N \cH_i \;   ,
\ee
which is a tensor product of partial Hilbert spaces
\be
\label{2}
  \cH_i = {\rm span}\{ | n_i \rgl \} 
\ee
that are closed linear envelopes over the related bases. So that the total Hilbert 
space can be written as
\be
\label{3}
 \cH = {\rm span} \left \{ \bigotimes_{i=1}^N  | n_i \rgl \right \} .
\ee
 
{\it Disentangled states} have the form of the products
\be
\label{4}
\vp_{dis} = \bigotimes_{j=1}^N \vp_j 
\ee
of the partial states  
\be
\label{5}
\vp_j = \sum_{n_j} c_{n_j} | n_j \rgl \in \cH_j
\ee
pertaining to the partial Hilbert spaces. Respectively, a disentangled state
is of the form
\be
\label{6}
 \vp_{dis} = \bigotimes_{j=1}^N \sum_{n_j}  c_{n_j} | n_j \rgl \; .
\ee
All disentangled states of the space $\mathcal{H}$ compose a {\it disentangled set}
\be
\label{7}
 \cD \equiv \{ \forall \vp_{dis} \in \cH \} \;  .
\ee

{\it Entangled states} of the space $\mathcal{H}$ can be written as
\be
\label{8}
 \vp_{ent} = \sum_{ \{ n_i\} } c_{n_1 n_2 \ldots n_N} |\; n_1 n_2 \ldots n_N \rgl \;  ,
\ee
where at least one of the coefficients cannot be represented as a product of $c_{n_i}$,
so that at least for one coefficient
\be
\label{9}
 c_{n_1 n_2 \ldots n_N} \neq \prod_{i=1}^N c_{n_i} \;  .
\ee
The collection of all entangled states forms an {\it entangled set}
\be
\label{10}
 \cH\setminus \cD \equiv \{ \vp_{ent} \in \cH , \; \vp_{ent} \not\in \cD \} \;  .
\ee

Thus the total Hilbert space $\mathcal{H}$ consists of two sets, a disentangled
set $\mathcal{D}$ and an entangled set $\mathcal{H} \setminus \mathcal{D}$.

\section{Measure of entanglement production}

One should distinguish entangled states and entanglement production by quantum 
operations. Entanglement of states can be induced by operators acting on the states
of a Hilbert space. So, entangled states characterize a static feature of a
considered set, while entanglement production describes the operational feature
of an operator.   

Let $\hat{A}$ be a non-singular trace-class operator on $\mathcal{H}$, such that
\be
\label{11}
 0\neq | {\rm Tr}_\cH \hat A | < \infty \;  .
\ee
The action of the operator on a disentangled state $\vp_{dis} \in \mathcal{D}$ 
can result either in a different disentangled state $\vp_{dis}' \in \mathcal{D}$ 
or in an entangled state $\vp_{ent}\in \mathcal{H} \setminus \mathcal{D}$. How 
would it be possible to quantify the entangling ability of an operator?

For two given states $\vp_{dis}\in\mathcal{D}$ and 
$\vp_{ent}\in\mathcal{H}\setminus\mathcal{D}$, one could define a kind of the 
entanglement probability
\be
\label{12}
 p_{ent}\left ( \hat A\vp_{dis} \ra \vp_{ent} \right ) \propto 
| ( \vp_{ent} | \hat A \vp_{dis} ) |^2  \; ,
\ee
being appropriately normalized. However, this would be not a general property of 
the operator on the whole Hilbert space, but merely its property with regard to the
two given particular states.  

The global ability of an operator of producing entanglement in the total Hilbert 
space $\mathcal{H}$, when
\be
\label{13}
 \hat A \cD \ra \cH \setminus \cD \;  ,
\ee
can be quantified as follows.   

The entangling ability of the operator $\hat{A}$ can be understood by comparing its 
action on the given Hilbert space with the action of its non-entangling counterpart
$\hat{A}^{\otimes}$ that does not entangle the states,
\be
\label{14}
 \hat A^\otimes \cD \ra \cD \; ,
\ee
being defined as
\be
\label{15}
  \hat A^\otimes \equiv 
\frac{\bigotimes_{i=1}^N \hat A_i}{({\rm Tr}_\cH \hat A)^{N-1}} \;  ,
\ee
where the partially traced operators
\be
\label{16}
\hat A_i \equiv {\rm Tr}_{\cH/\cH_i} \hat A
\ee
are obtained by tracing out the operator $\hat{A}$ over all partial Hilbert spaces,
composing $\mathcal{H}$, except one space $\mathcal{H}_i$. The denominator
in the non-entangling counterpart is such that the normalization
\be
\label{17}
{\rm Tr}_\cH \hat A^\otimes = {\rm Tr}_\cH \hat A
\ee
be preserved. 

The measure of entanglement production, for an operator $\hat{A}$, is defined 
\cite{Yukalov_8,Yukalov_9} as
\be
\label{18}
 \ep(\hat A) \equiv 
\log \; \frac{||\hat A||_p}{||\hat A^\otimes ||_p} \;  .
\ee
The logarithm can be taken with respect to any base. The norm here can be accepted
as a Schatten $p$-norm
$$
 ||\hat A||_p \equiv \left ( {\rm Tr}_\cH |\; \hat A \; |^p \right )^{1/p} \;  ,
$$
in which
$$
p \in [ 1, \infty ) \; , \qquad |\; \hat A \; | \equiv \sqrt{\hat A^+ \hat A} \;  .
$$

The Schatten norms are isometrically invariant, and therefore unitary invariant, 
such that
$$
 ||\hat A||_p = || U_1 \hat A U_2 ||_p
$$
for any linear isometries, hence, for any unitary transformations $U_1$ and $U_2$. 
Thence, the Schatten norms are independent of the basis used when taking the 
trace operation.

It is convenient to employ the Hilbert-Schmidt, or Frobenius, norm
\be
\label{19}
  ||\hat A||_2 = \sqrt{ {\rm Tr}_\cH ( \hat A^+ \hat A ) } \; ,
\ee
which is the Schatten $2$-norm. 

The standard operator norm corresponds to the Schatten $\infty$-norm
\be
\label{20}
  ||\hat A||_\infty \equiv \sup_{\vp\in\cH} \; \frac{||\hat A\vp||}{||\vp||}
\qquad ( \vp \neq 0 ) \;  ,
\ee
denoted in what follows as
$$
  ||\hat A|| \equiv  ||\hat A||_\infty \;  .
$$

Keeping in mind an orthonormal basis $\{e_\alpha\}$ in $\mathcal{H}$, we can write
$$
 {\rm Tr}_\cH ( \hat A^+ \hat A) = \sum_\al (e_\al , \hat A^+ \hat A e_\al ) =
\sum_\al ( \hat A e_\al , \hat A e_\al ) = \sum_\al || \hat A e_\al ||^2 \;  .
$$
Therefore the Hilbert-Schmidt norm can be defined as
$$
 ||\hat A||_2^2 = \sum_\al ||\hat A e_\al ||^2 \;   .
$$

Quantity (\ref{18}) enjoys all necessary properties required for being a measure.
Thus, it is semi-positive. Because of the importance of this property, we prove 
it below.  

\vskip 2mm

{\bf Theorem}. For trace-class operators $\hat{A}: \mathcal{H} \ra \mathcal{H}$
and $\hat{A}^\otimes: \mathcal{D} \ra \mathcal{D}$, defined in (\ref{15}), the 
measure of entanglement production (\ref{18}) is semi-positive,
\be
\label{21}
\ep ( \hat A) \geq 0 \; .
\ee

\vskip 2mm

{\it Proof}. The operators, enjoying finite norm (\ref{19}), are Hilbert-Schmidt
operators. Note that the trace-class operators are also Hilbert-Schmidt operators.
A family of the operators $\hat{A}: \mathcal{H} \ra \mathcal{H}$, supplemented 
by norm (\ref{19}), forms a Banach space
\be
\label{22}
 \cB(\cH ) \equiv \{ \hat A : \; \cH \ra \cH , || \hat A ||_2 \} \; ,
\ee
that is a normed, complete, linear space. Respectively, a family of the operators 
$\hat{A}^\otimes: \mathcal{D} \ra \mathcal{D}$, complimented by norm (\ref{19}), 
forms a Banach space
\be
\label{23}
 \cB_\otimes(\cD ) \equiv 
\{ \hat A^\otimes : \; \cD \ra \cD , || \hat A^\otimes ||_2 \} \;  ,
\ee
which is a subspace of space (\ref{22}),
$$
 \cB_\otimes(\cD ) \subset \cB(\cH) \;  .
$$
It is possible to introduce a projection $P_\otimes$ in the space of bounded 
linear operators \cite{Kuo_10}, such that
\be
\label{24}
  \cP_\otimes \cB(\cH ) = \cB_\otimes(\cD ) \; ,
\ee
with the standard properties of idempotence and self-adjointness,
$$
\cP_\otimes^2 = \cP_\otimes \; , \qquad 
\cP_\otimes^+ = \cP_\otimes \;   .
$$
The operator norm of the projector reads as
$$
 || \cP_\otimes || = \sup_{\hat A \in \cB(\cH)} \; 
\frac{|| \cP_\otimes \hat A||}{|| \hat A|| } \qquad ( \hat A \neq 0 ) \; ,
$$
where in the right-hand side the operator norms are given by (\ref{20}). Then
$$
 || \cP_\otimes || = \sup_{\hat A \in \cB(\cH)} \; \sup_{\vp\in \cH} \;
\frac{ || \cP_\otimes \hat A \vp ||}{|| \hat A \vp || } \qquad
(\hat A \neq 0 \; , \; \vp \neq 0 ) \;  ,
$$
expressed through the vector norms. Since $\hat{A} \varphi \in \mathcal{H}$, 
the projector norm takes the form
$$
 || \cP_\otimes || = \sup_{\hat A \vp \in \cH} \; 
\frac{ || \cP_\otimes \hat A \vp ||}{|| \hat A \vp || } \qquad 
( \hat A \vp \neq 0 ) \; .
$$

The non-entangling counterpart (\ref{15}) is obtained from the initial operator 
$\hat{A}$ by means of the transformation
$$
 \cP_\otimes \hat A = 
\hat A^\otimes \in \cB_\otimes(\cD) \subset \cB(\cH)  
$$
preserving normalization (\ref{17}). Therefore
$$
|| \cP_\otimes || =  1 \;  .
$$ 
Keeping in mind that 
$$
 || \hat A^\otimes ||_2^2 = ||  \cP_\otimes \hat A||_2^2 = 
\sum_\al || \cP_\otimes \hat A e_\al ||^2 \; ,
$$
we have
$$
  || \cP_\otimes \hat A e_\al || \leq 
|| \cP_\otimes ||\; || \hat A e_\al ||  \;  .
$$
Hence
$$
\sum_\al || \cP_\otimes \hat A e_\al ||^2 \leq \sum_\al || \hat A e_\al ||^2
= || \hat A ||_2^2 \;  ,
$$ 
from where
\be
\label{25}
 || \hat A^\otimes ||_2 \leq  || \hat A ||_2 \; ,
\ee
which implies that measure (\ref{18}) is semi-positive, satisfying (\ref{21}). $\square$

\vskip 3mm
Employing the properties of the Hilbert-Schmidt norm \cite{Reed_11,Horn_12},
it is straightforward to prove that measure (\ref{18}) satisfies all other conditions
required for being a measure \cite{Yukalov_9}. Thus, it is continuous with 
respect to the norm convergence, it is zero for non-entangling operators, it is 
additive with respect to copies, and invariant under local unitary operations.

\section{Evolutional entanglement production}

Entanglement production has been considered for density matrices of two-mode 
Bose systems of trapped atoms and multimode Bose systems in optical lattices
\cite{Yukalov_13,Yukalov_14,Yukalov_15,Yukalov_16,Yukalov_17,Yukalov_18,Yukalov_19,Yukalov_20}.
Also, it has been studied for pseudospin entanglement production of radiating 
systems \cite{Yukalov_21}. Here we investigate the entanglement production by the
evolution operators \cite{Yukalov_22}.

The evolution operator of a closed quantum system,
\be
\label{26}
 \hat U(t) = e^{-i Ht} \;  ,
\ee
characterizes the evolution of quantum states in time, generated by a Hamiltonian
$H$. Even when an initial state $\varphi(0)$ would be not entangled, it can become 
entangled with time yielding an entangled state
\be
\label{27}
 \vp(t) = \hat U(t) \vp(0) \;  .
\ee
The entangling properties of the evolution operator are described by the 
entanglement-production measure
\be
\label{28}
 \ep(\hat U(t) ) = \log \; \frac{||\hat U(t)||_2}{||\hat U^\otimes(t)||_2}
\equiv \ep(t) \;  .
\ee

Since evolution operators are unitary, so that
$$
 \hat U^+(t) \hat U(t) = \hat 1_\cH \;  ,
$$
with the right-hand side being the identity operator in $\mathcal{H}$, the 
related norm equals the dimensionality of the Hilbert space $\mathcal{H}$,  
\be
\label{29}
 ||\hat U(t)||_2^2 = {\rm Tr}_\cH \hat 1_\cH = {\rm dim} \cH \;  .
\ee

The calculation of the norm for the nonentangling counterpart $||\hat{U}^\otimes||_2$
is more involved. In the limit of short times, we have
$$
||\hat U^\otimes(t)||_2^2 \simeq {\rm dim} \cH - \mu t^2 \;  ,
$$
with a constant $\mu$ depending on the system Hamiltonian. As a result, the 
short-time behavior of the entanglement-production measure is
\be
\label{30}
  \ep(t) \simeq \frac{1}{2} \; \mu t^2 \qquad (t\ra 0 ) \;  .
\ee

\section{Entanglement for two-qubit registers}

Two-qubit registers are widely used for quantum information processing 
\cite{Williams_1,Nielsen_2,Vedral_3,Keyl_4,YS_5,Wilde_6,YS_7}. Such registers 
can be realized with the help of cold atoms, trapped ions, spin systems, and 
photons in cavities 
\cite{Yukalov_19,Gallagher_23,Di_24,Raithel_25,Choi_26,Gallagher_27,Saffman_28,Buluta_29,Murray_30}.
Below we show how the entanglement-production measure for arbitrary time can 
be explicitly calculated for two-qubit registers. 

In the case of a two-qubit register, a disentangled state is a product
$$
\vp_{dis} = \vp_1 \bigotimes \vp_2 \;   ,
$$
in which
$$
 \vp_1 = a_1 | \uparrow \rgl + a_2 | \downarrow \rgl \; , \qquad
 \vp_2 = b_1 | \uparrow \rgl + b_2 | \downarrow \rgl \; .
$$
And an entangled state reads as
$$
 \vp_{ent} = c_1 | \uparrow \uparrow \rgl + c_2 | \uparrow \downarrow \rgl +
 c_3 | \downarrow \uparrow \rgl + c_4 | \downarrow \downarrow \rgl \;  ,
$$
where at least one of the coefficients $c_j$ is not a product of the type $a_j b_j$. 
 
The two-qubit Hamiltonian has the form
\be
\label{31}
 H = H_0 + H_{int} \;  ,
\ee
consisting of a free term  
\be
\label{32}
  H_0 = - h \left ( S_i^z \bigotimes \hat 1_2 \; + \;
\hat 1_1 \bigotimes S_2^z \right ) \; ,
\ee
where $S_j^z$ are spin operators, while $h$ is an external field, and the 
interaction term is
\be
\label{33}  
 H_{int} = 2J S_1^z \bigotimes S_2^z \;  .
\ee
The Hamiltonian is of a two-site Ising type with the interaction strength $J$.

Calculating the entanglement-production measure, as explained in the previous 
sections, we obtain
\be
\label{34}
\ep(t) = \frac{1}{2} \; 
\log \; \frac{1+\cos^2(ht) + 2\cos(ht)\cos(Jt)}{[1+\cos(ht)\cos(Jt)]^2} \; .
\ee
At the initial time, as it should be, no entanglement is yet produced by the 
evolution operator, 
$$
 \lim_{t\ra 0} \ep(t) = \ep(0) = 0 \; .
$$ 
Also, notice the importance of interactions, without which no entanglement 
could be produced,
$$
 \lim_{J\ra 0} \ep(t) = 0 \qquad ( t \geq 0) \;  .
$$
The measure is invariant with respect to the inversion of the signs in $h$ and $J$.  
At short times, the measure behaves as
\be
\label{35}
 \ep(t) \simeq \frac{J^2}{8} \; t^2 + \frac{J^2(J^2-12h^2)}{192} \; t^4 \; ,
\ee
in agreement with formula (\ref{30}).   

The temporal behavior of the measure is either periodic, when the ratio $h/J$
is rational, or quasi-periodic, if this ratio is irrational. This is illustrated
by Figs. 1 and 2, where time is measured in units of $1/J$.

\section{Conclusion and possible extensions}

There exist two types of entanglement measures that characterize either the 
entanglement of given states or the amount of entanglement produced by quantum
operations. Here the latter notion is considered quantified by a measure of 
entanglement production. The theorem on the semi-positiveness of this measure
is proved. The entanglement produced by evolution operators is analyzed. 
Detailed calculations are accomplished for a two-qubit register modeled by 
an Ising-type Hamiltonian. The measure of entanglement production for such a 
case is either periodic or quasi-periodic in time, depending on the relation 
between the values of an external field and pseudospin interactions.   

It is straightforward to notice that in the same way, as has been done in the 
present paper, one can define {\it thermal entanglement production} by statistical
operators of equilibrium systems. Let $\hat{\rho}$ be a statistical operator.
Since this operator is trace-normalized to one, its non-entangling, or distilled, 
counterpart reads as
$$
 \hat\rho_\otimes = \bigotimes_{i=1}^N \hat\rho_i \;  ,
$$
with the partial statistical operators
$$  
\hat\rho_i \equiv {\rm Tr}_{\cH/\cH_i} \hat\rho \;   .
$$
The measure of entanglement production by the statistical operator is  
$$
 \ep(\hat\rho) = \log \; \frac{||\hat\rho||_2}{||\hat\rho_\otimes||_2} \;  .
$$

In equilibrium, the statistical operator of a system with a Hamiltonian $H$ 
has the form
$$
 \hat\rho = \frac{1}{Z} \; e^{-\bt H} \; , \qquad 
Z = {\rm Tr}_\cH e^{-\bt H} \;  ,
$$
where $\beta = 1/ T$ is inverse temperature. It easy to notice that the exponential 
here can be represented through the evolution operator in imaginary time as
$$
e^{-\bt H} = \hat U(-i\bt) \;   .
$$
Hence we have
$$
 \hat\rho = \frac{1}{Z} \; \hat U(- i\bt) \; , \qquad
Z = {\rm Tr}_\cH \hat U(-i\bt) \;  .
$$
With the partial term 
$$
 \hat\rho_i = \frac{1}{Z} \; {\rm Tr}_{\cH/\cH_i} \hat U(-i\bt) \;  ,
$$
the distilled operator becomes
$$
\hat\rho_\otimes = 
\frac{1}{Z^N} \bigotimes_{i=1}^N {\rm Tr}_{\cH/\cH_i} \hat U(-i\bt) \; .
$$
Thus we come to the measure
$$
 \ep(\hat\rho) = \frac{1}{2} \; \log \; 
\frac{Z^{2N-2}{\rm Tr}_\cH \hat U(-2i\bt)}
{\prod_{i=1}^N{\rm Tr}_{\cH_i} [ {\rm Tr}_{\cH/\cH_i}\hat U(-i\bt)]^2} \;  .
$$
The calculational procedure for this measure, as is seen, reduces to that 
for the evolution operator. A detailed investigation of the thermal 
entanglement production, based on the above measure, will be treated in a 
separate paper.

\section*{Acknowledgement}

Financial support from the Russian Foundation for Basic Research
(grant $\#$ 14-02-00723) is appreciated.

\vskip 5mm

\begin{figure}[ht]
\centerline{
\hbox{ \includegraphics[width=6.7cm]{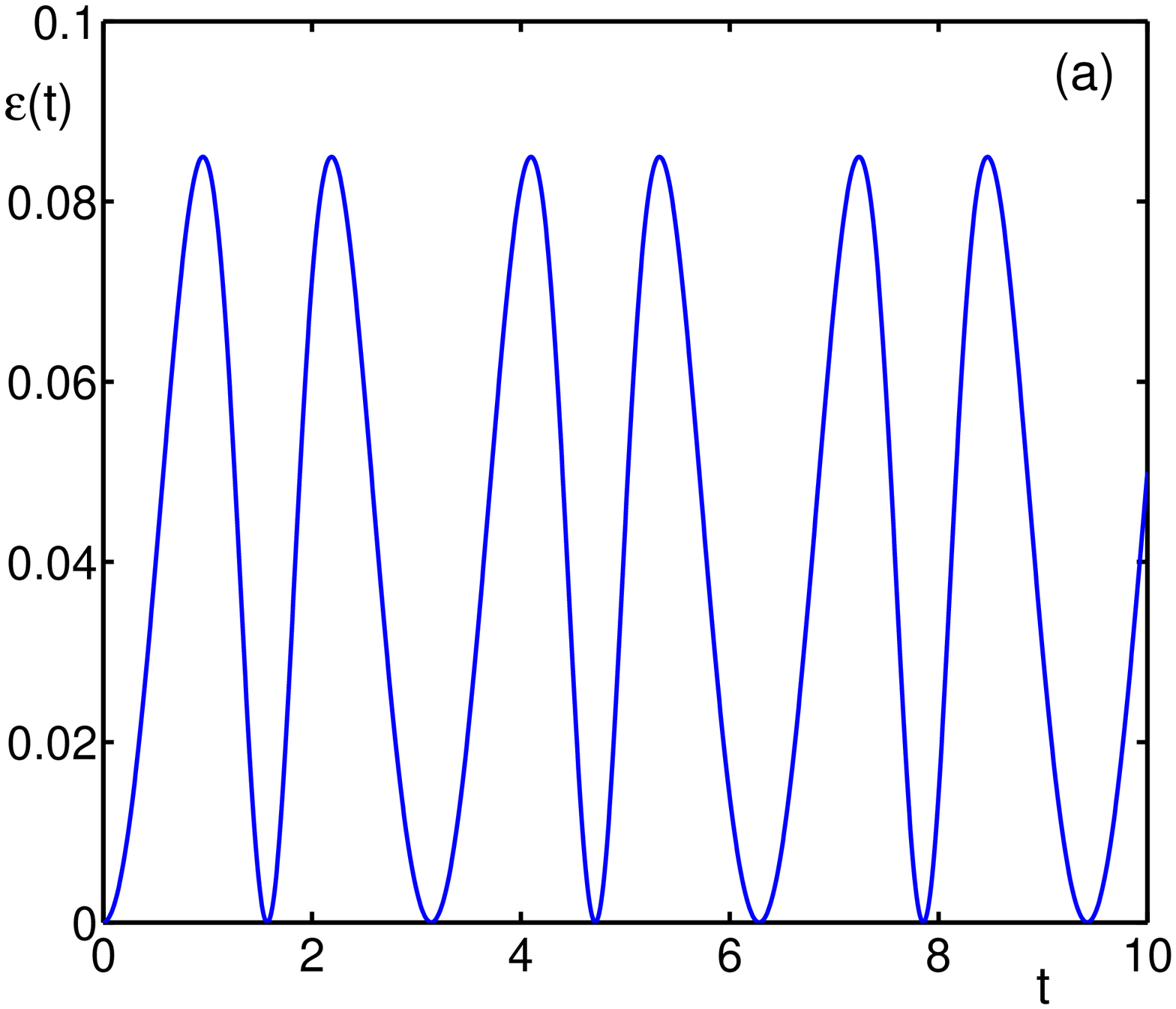} \hspace{1cm}
\includegraphics[width=6.7cm]{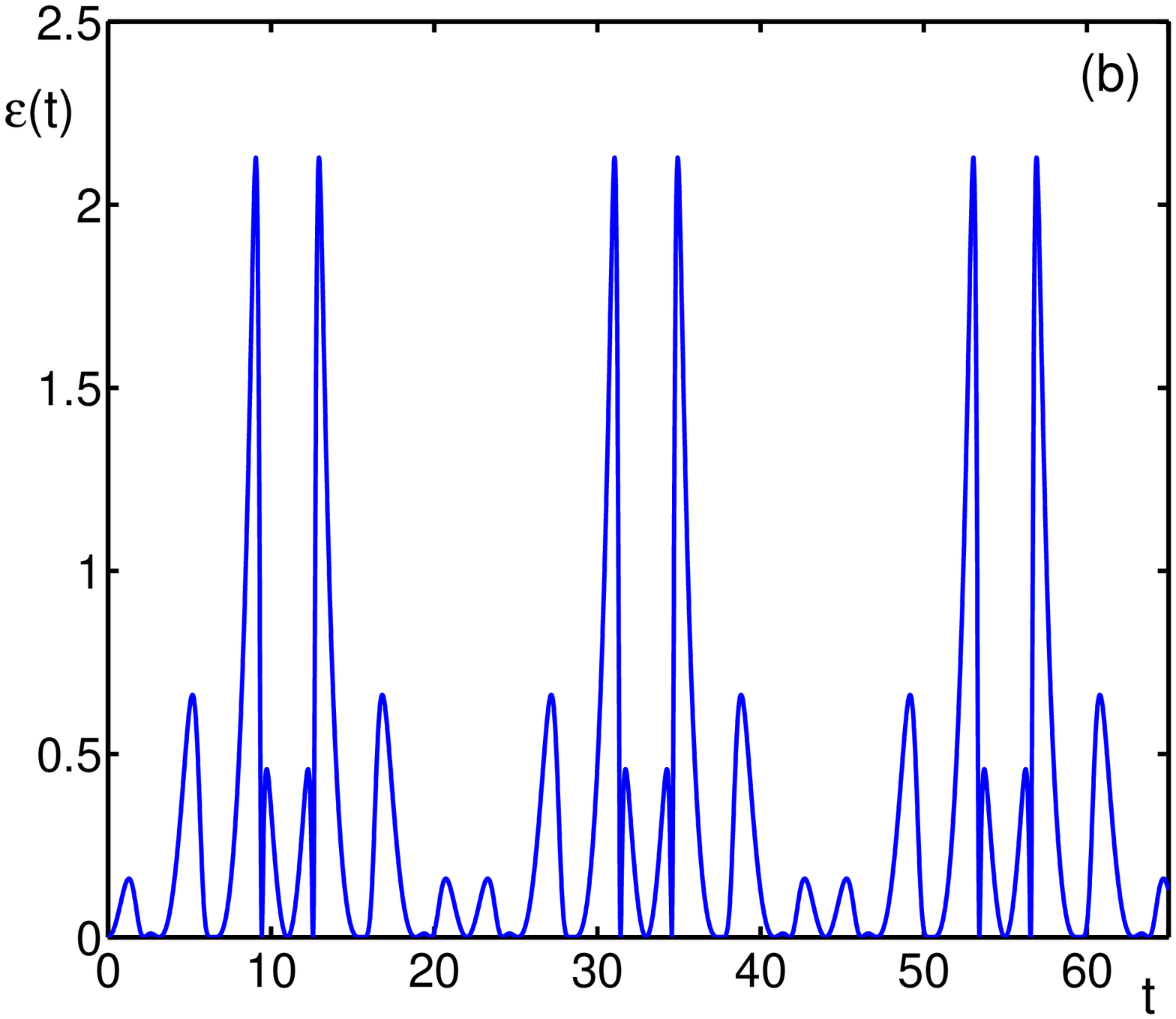}  } }
\vspace{10pt}
\centerline{
\hbox{ \includegraphics[width=6.7cm]{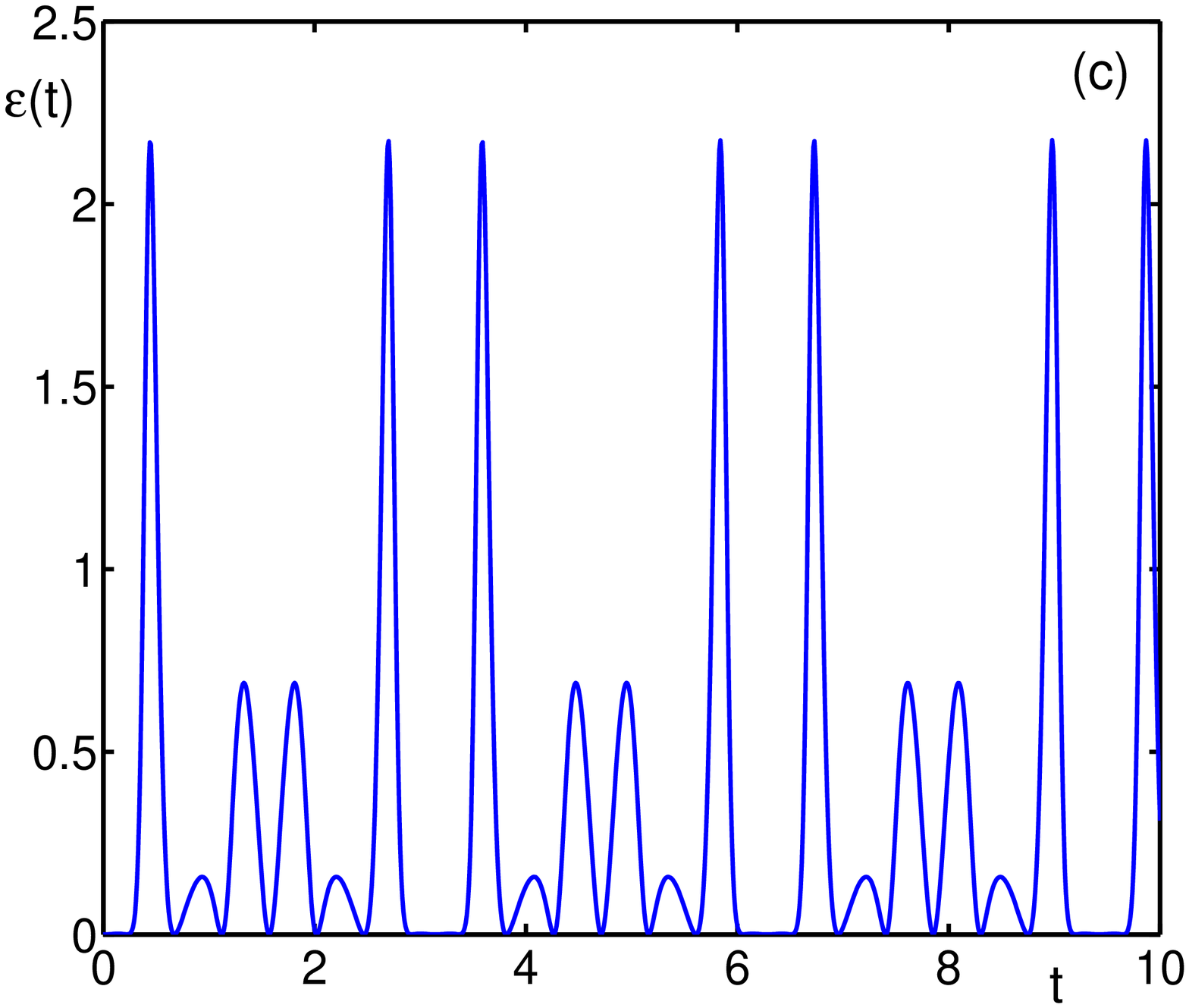} \hspace{1cm}
\includegraphics[width=6.7cm]{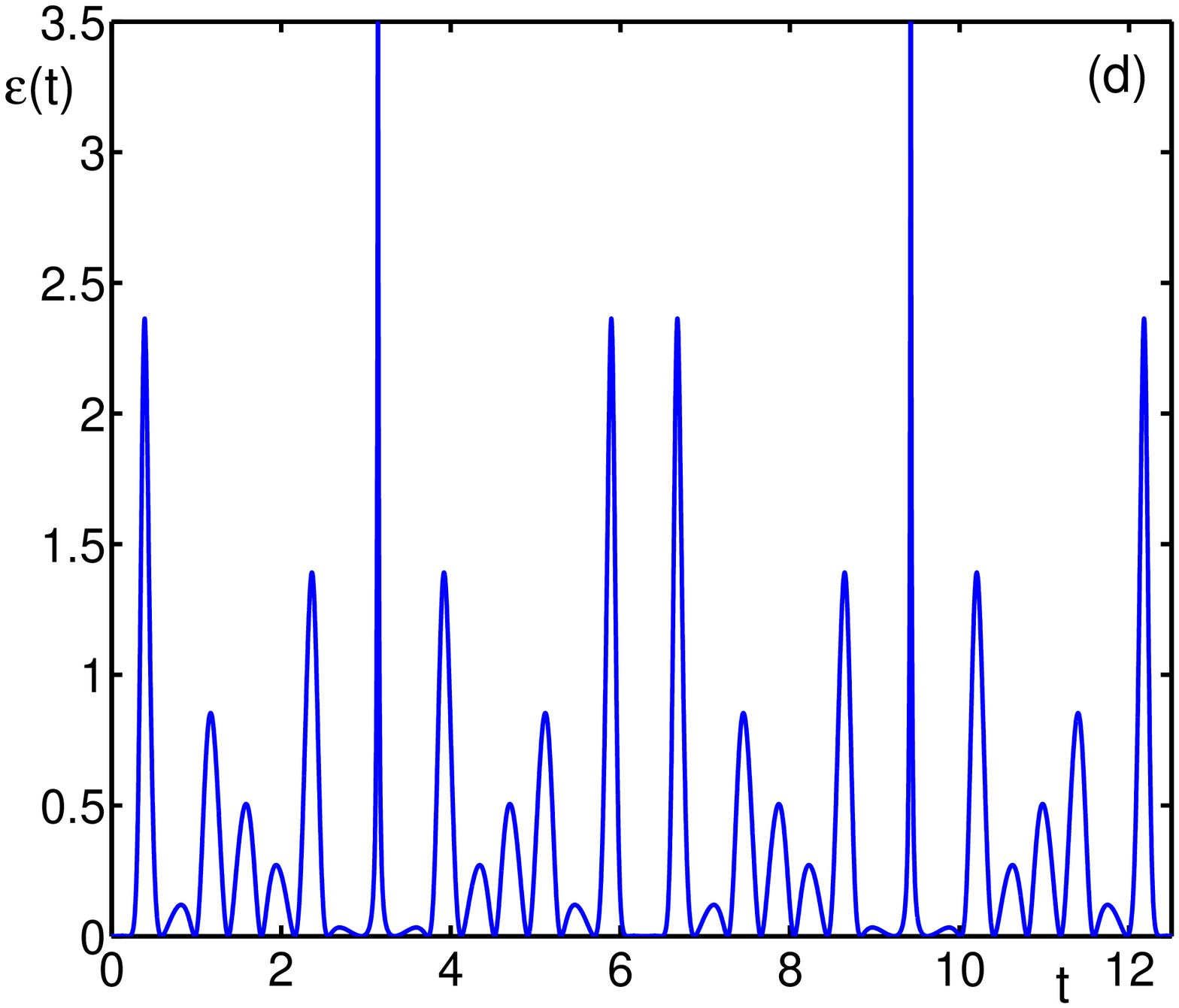} } }
\caption{The entanglement-production measure, for the case of periodic
evolution, as a function of time measured in units of $1/J$, for different
fields: (a) $h/J = 1$ (the period is $\pi$); (b) $h/J = 5/7$ (the period
is $7\pi$); (c) $h/J = 7$ (the period is $\pi$); (d) $h/J = 8$ (the period
is $2\pi$).
}
\label{fig:Fig.1}
\end{figure}

\newpage

\begin{figure}[ht]
\centerline{
\hbox{ \includegraphics[width=6.7cm]{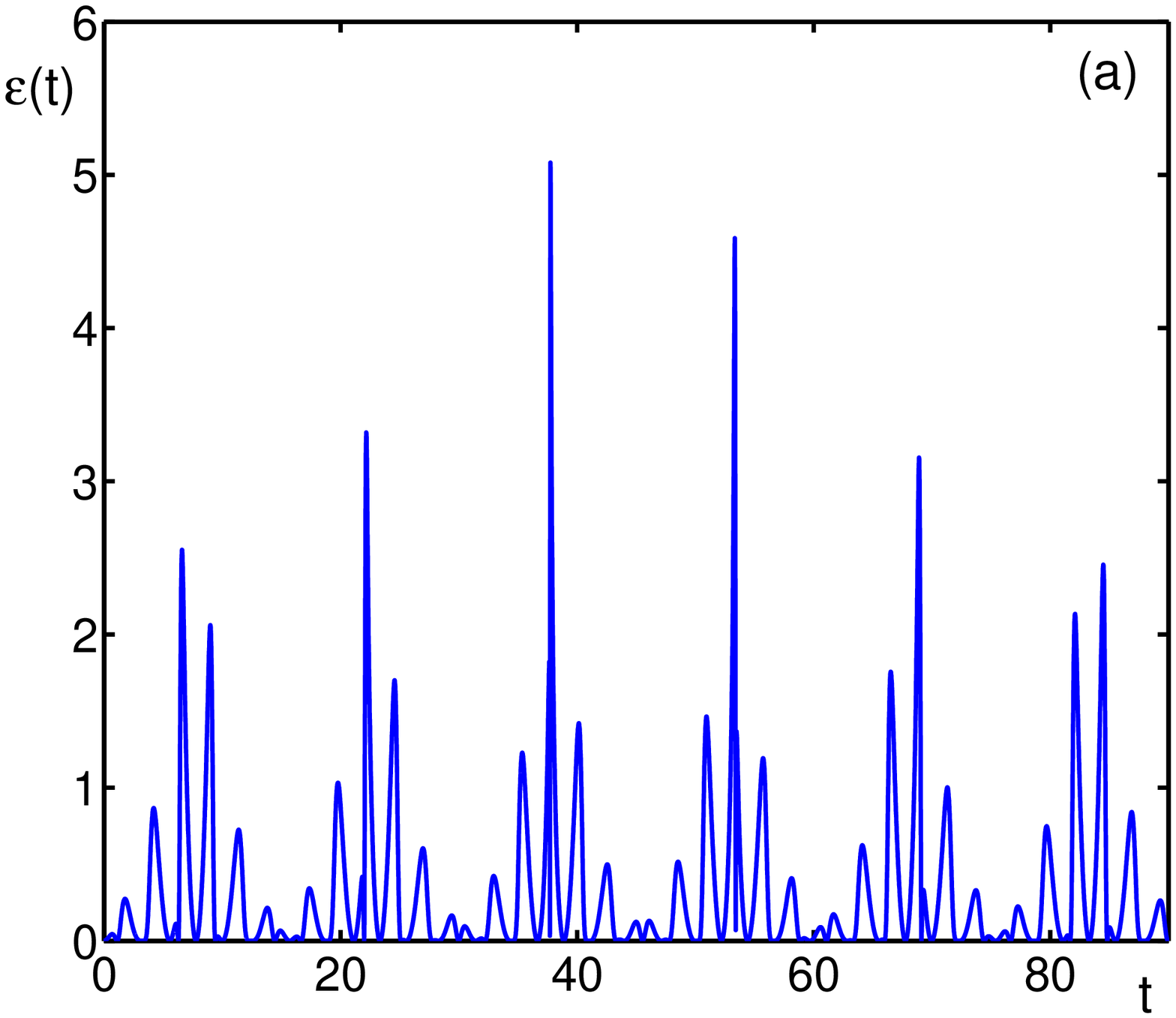} \hspace{1cm}
\includegraphics[width=6.7cm]{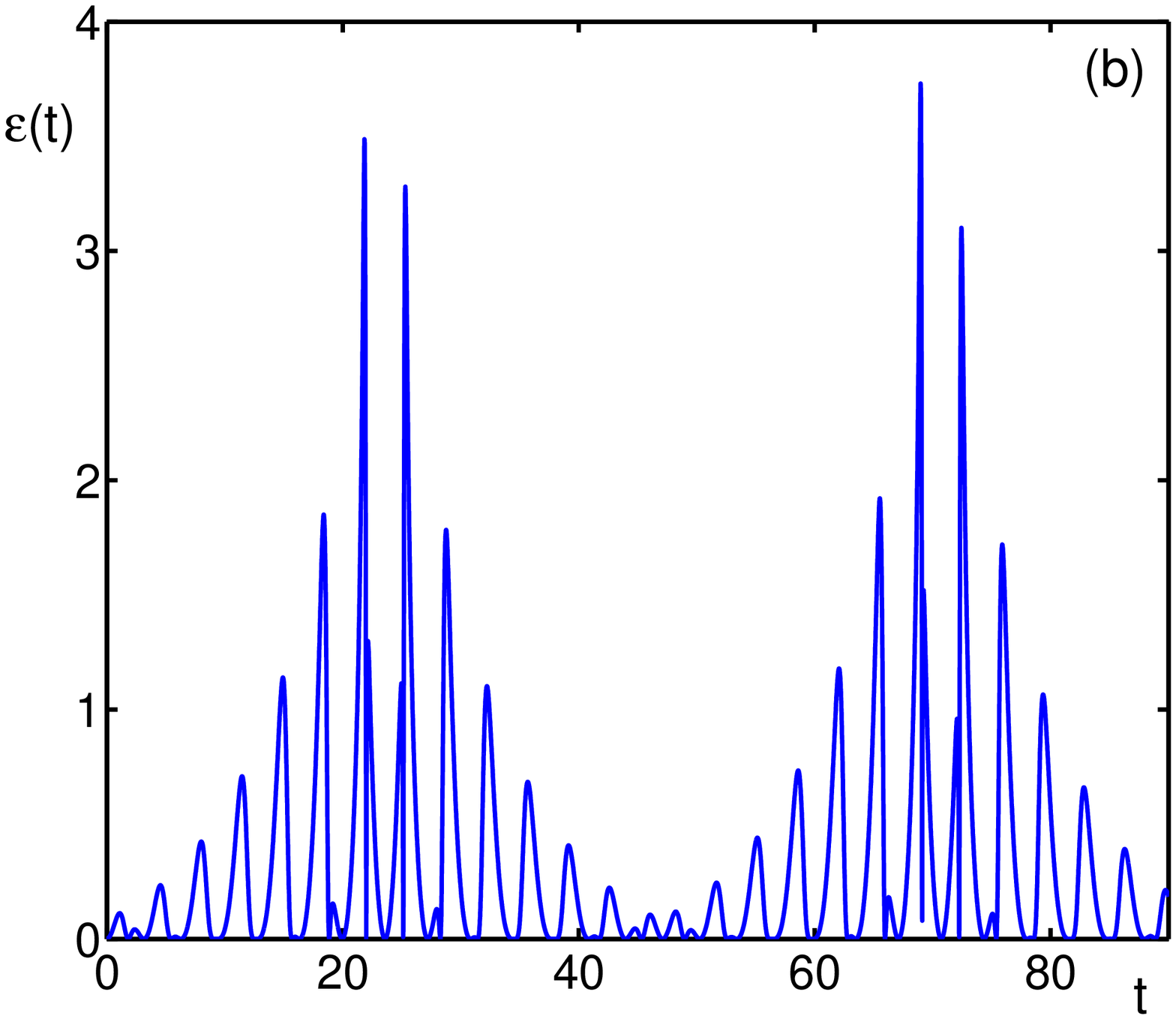}  } }
\vspace{10pt}
\centerline{
\hbox{ \includegraphics[width=6.7cm]{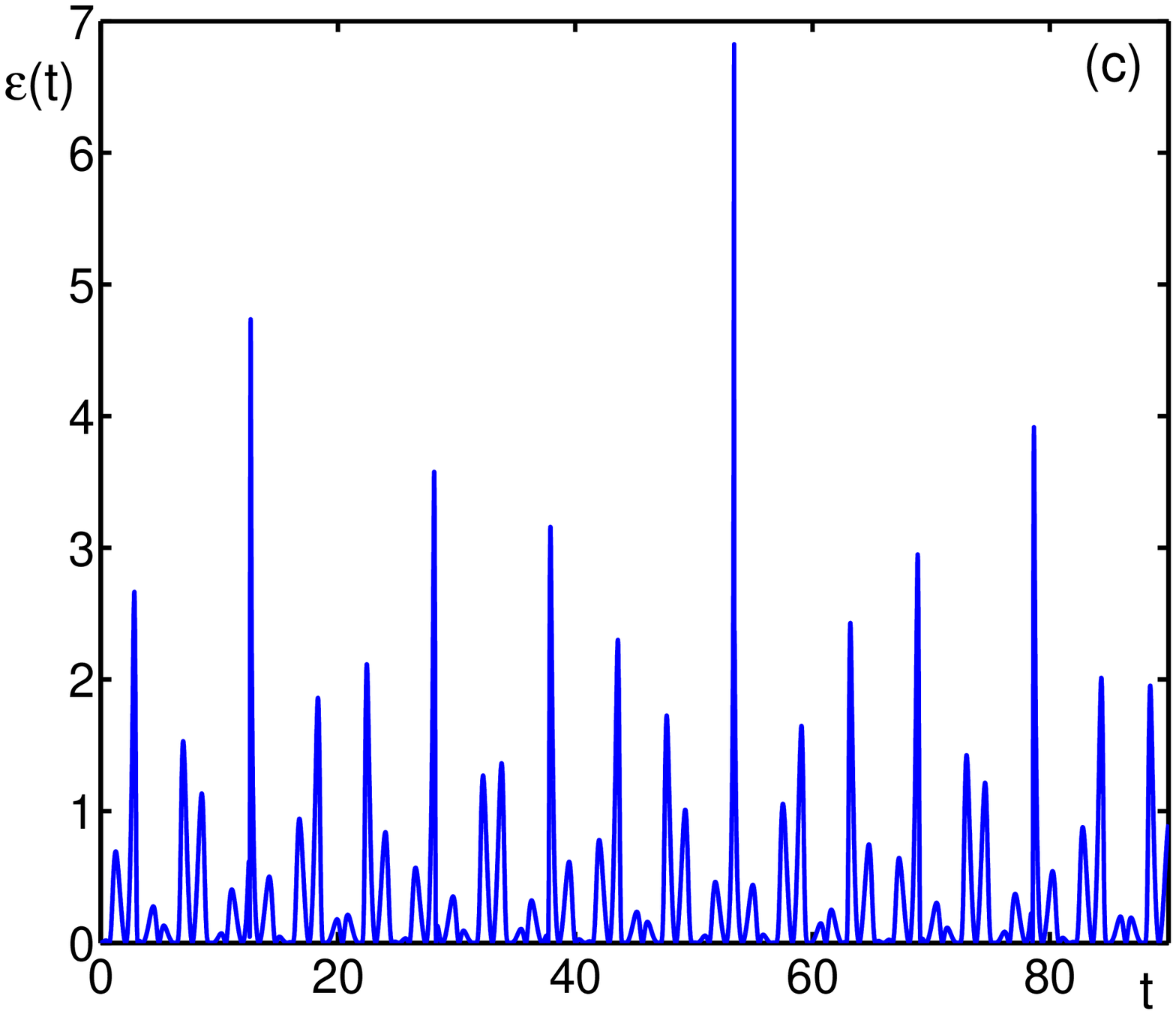} \hspace{1cm}
\includegraphics[width=6.7cm]{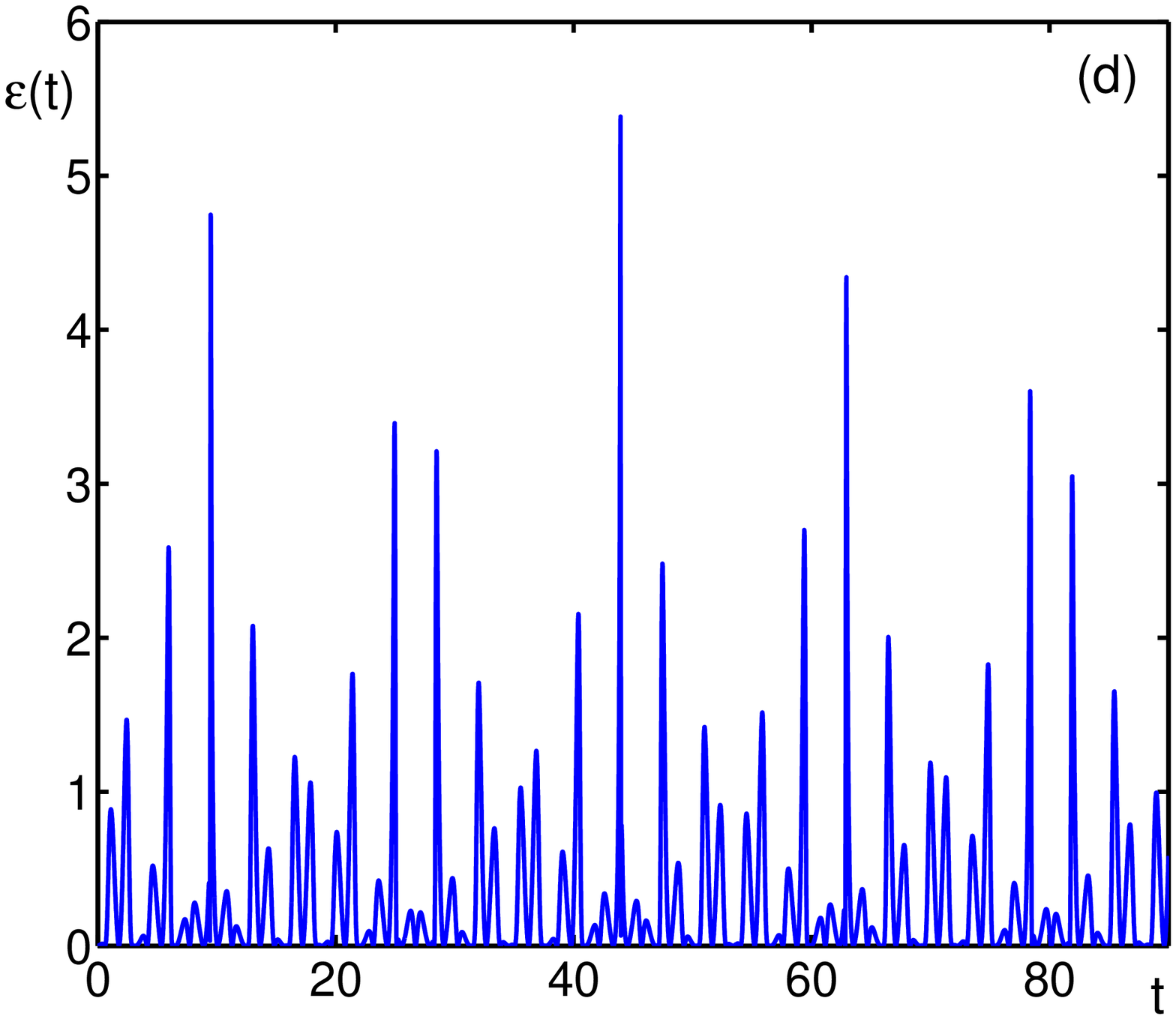} } }
\caption{The measure of evolutional entanglement production, illustrating
quasi-periodic behavior, for different fields: (a) $h/J = \sqrt{2}$;
(b) $h/J = \sqrt{3}/2$; (c) $h/J = \sqrt{5}$; (d) $h/J = \sqrt{7}$.
}
\label{fig:Fig.2}
\end{figure}

\end{document}